\documentclass[
preprint,
showpacs,
showkeys,
 amsmath,amssymb,
 aps,
floatfix,
]{revtex4-1}
\nonstopmode
\usepackage{graphicx}% Include figure files
\usepackage{dcolumn}% Align table columns on decimal point
\usepackage{bm}% bold math
\begin{document}

%\preprint{APS/123-QED}

\title{Universal Constituent-Quark Model for Baryons}% Force line breaks with \\

\author{Joseph P. Day}
\author{Willibald Plessas}
\affiliation{%
Institute of Physics, University of Graz, A-8010 Graz, Austria
}%
\author{Ki-Seok Choi}%
\affiliation{
Department of Physics, Soongsil University, Seoul 156-743, Republic of Korea
}%


%\affiliation{
%\affiliation{%


\date{\today}% It is always \today, today,

\begin{abstract}
We present a relativistic constituent-quark model that covers all known baryons
from the nucleon up to $\Omega_{bbb}$. The corresponding invariant mass operator includes a linear confinement and a hyperfine interaction based on effective degrees of freedom.
The model provides for a unified description of practically all baryon spectra in good
agreement with present phenomenology and it can tentatively be employed for the relativistic treatment of all kinds of baryon reactions. Predictions of states still missing
in the phenomenological data base, especially in the lesser explored heavy-flavor sectors of
charm and bottom baryons, should be important especially for future experiments in these areas.
\end{abstract}

%Use showkeys class option if keyword
\maketitle

%\tableofcontents

%\section{Introduction}

Quantum chromodynamics (QCD) is generally considered as the fundamental theory of strong interactions. While it has been well established in the perturbative regime at high energies,
QCD still lacks a comprehensive solution at low and intermediate energies, even 40 years
after its invention. In order to deal with the wealth of non-perturbative phenomena, various
approaches are followed with limited validity and applicability. This is especially also true
for lattice QCD, various functional methods, or chiral perturbation theory, to name only
a few. In neither one of these approaches the full dynamical content of QCD can yet be included.
Basically, the difficulties are associated with a relativistically covariant treatment
of confinement and the spontaneous breaking of chiral symmetry (SB$\chi$S), the latter
being a well-established property of QCD at low and intermediate energies.
As a result, most hadron reactions, like
resonance excitations, strong and electroweak decays etc., are nowadays only amenable
to models of QCD. Most famous is the constituent-quark model (CQM), which essentially relies
on a limited number of effective degrees of freedom with the aim of encoding the essential
features of low- and intermediate-energy QCD.

The CQM has a long history, and it has made important contributions to the understanding of
many hadron properties, think only of the fact that the systematization of hadrons in the
standard particle-data base~\cite{Nakamura:2010zzi} follows the valence-quark picture.
Over the decades the CQM has ripened into a stage where its formulation and solution are
well based on a relativistic (or more generally Poincar\'e-invariant) quantum theory. Relativistic
constituent-quark models (RCQM) have been developed by several groups, however, with
limited domains of validity. Of course, it is desirable to have a framework as universal
as possible for the description of all kinds of hadron processes in the low- and
intermediate-energy regions. This is especially true in view of the advent of ever
more data on heavy-baryon spectroscopy from present and future experimental facilities.

Here, we present a RCQM that comprises all known baryons with flavors $u$, $d$, $s$,
$c$, and $b$ within a single framework. There have been only a few efforts so far
to extend a CQM from light- to heavy-flavor baryons. We may mention, for example, the
approach by the Bonn group who have developed a RCQM, based on the 't Hooft instanton
interaction, along a microscopic theory solving the Salpeter equation~\cite{Loring:2001kx}
and extended their model to charmed baryons~\cite{Migura:2006ep}, still not yet
covering bottom baryons. A further quark-model
attempt has been undertaken by the Mons-Li\`ege group relying on the large-$N_c$
expansion~\cite{Semay:2007cv,Semay:2007ff}, partially extended to heavy-flavor
baryons~\cite{Semay:2008wn}. Similarly, efforts are invested to expand other approaches
to heavy baryons, such as the employment of Dyson-Schwinger equations together with either
quark-diquark or three-quark calculations~\cite{SanchisAlepuz:2011aa,SanchisAlepuz:2012}.
Also an increased amount of more refined lattice-QCD results has by now become available
on heavy-baryon spectra (see, e.g., the recent work by Liu et al.~\cite{Liu:2009jc} and
references cited therein).

%\section{Model}

Our RCQM is based on the invariant mass operator
\begin{equation}
\hat{M}=\hat{M}_{\text{free}} + \hat{M}_{\text{int}} \;,
\label{massop}
\end{equation}
where the free part corresponds to the total kinetic energy of the three-quark system and
the interaction part contains the dynamics of the constituent quarks $Q$. In the rest frame
of the baryon, where its three-momentum
${\vec P}=\sum^3_i \vec k^2_i=0$, we may express the terms as
\begin{equation}
\hat M_{\text{free}}=\sum^3_{i=1}\sqrt{\hat m^2_i+\hat{\vec k}_i^2} \,,
\label{massopparts1}
\end{equation}
\begin{equation}
\hat M_{\text{int}}=\sum^3_{i<j}\hat V_{ij}=
\sum^3_{i<j}(\hat V^{\text{conf}}_{ij}+\hat V^{\text{hf}}_{ij}) \;.
\label{massopparts2}
\end{equation}
Here, the $\hat{\vec k}_i$ correspond to the three-momentum operators of the
individual quarks with
rest masses $m_i$ and the $Q$-$Q$ potentials $\hat V_{ij}$ are composed of confinement and
hyperfine interactions. By employing such a mass operator $\hat M^2= \hat P^\mu \hat P_\mu$\,, with baryon four-momentum $\hat P_\mu=(\hat H, \hat P_1, \hat P_2, \hat P_3)$, the Poincar\'{e} algebra involving all ten generators $\{\hat H, \hat P_i, \hat J_i, \hat K_i\}$,
 ($i=1,2,3$), or equivalently $\{\hat P_\mu, \hat J_{\mu\nu}\}$, ($\mu,\nu=0,1,2,3$),
of time and space translations, spatial rotations as well as Lorentz boosts, can be guaranteed. The solution of the eigenvalue problem of the mass operator $\hat M$ yields the
relativistically invariant mass spectra as well as the baryon eigenstates (the latter,
of course, initially in the standard rest frame).

We adopt the confinement depending linearly on the $Q$-$Q$ distance $r_{ij}$
\begin{equation}
V^{\text{conf}}_{ij}(\vec r_{ij})=V_0 +Cr_{ij}
\end{equation}
with the strength $C=2.33$ fm$^{-2}$, corresponding to the string tension of QCD.
The parameter $V_0=-402$ MeV is only necessary to set the ground state of the whole
baryon spectrum, i.e., the proton mass; it is irrelevant, if one considers only level
spacings.

The hyperfine interaction is most essential to describe all of the baryon excitation spectra.
In a unified model the hyperfine potential must be explicitly flavor-dependent. Otherwise,
e.g., the $N$ and $\Lambda$ spectra with their distinct level orderings could not be
reproduced simultaneously. At least for baryons with flavors $u$, $d$, and $s$ the type
of hyperfine interaction taking into account SB$\chi$S has been most successful over the
past years. Obviously, it grabs the essential degrees of freedom governing the behavior of
low-energy baryons~\cite{Manohar:1983md,Glozman:1995fu,Weinberg:2010bq}.
The RCQM constructed along this dynamical concept, i.e., on Goldstone-boson exchange (GBE),
has provided a comprehensive description
of all light and strange baryons~\cite{Glozman:1997ag,Glozman:1997fs}. This is not only true
with regard to the spectroscopy but to a large extent also for other baryon properties, like electromagnetic and axial form factors~\cite{Boffi:2001zb} and a number of other
reaction observables (for a concise summary see ref.~\cite{Plessas:2010pk}).
It has been tempting
to extend this successful concept even to the heavier flavors $c$ and $b$. By such studies
one should in addition learn about the proper light-heavy and heavy-heavy hyperfine $Q$-$Q$
interactions. Some exploratory work in this direction had already been undertaken some time
ago in ref.~\cite{Glozman:1995xy}, hinting to promising results also for charm and bottom
baryons.

Therefore we have advocated for the hyperfine interaction of our universal RCQM 
the $SU(5)_F$ GBE potential
\begin{equation}
V_{\text{hf}}(\vec{r}_{ij})=\bigg[ V_{24}(\vec{r}_{ij}) \sum_{a=1}^{24} \lambda_i^a \lambda_j^a
+ V_{0}(\vec{r}_{ij}) \lambda_i^0 \lambda_j^0  \bigg]\vec{\sigma}_i \cdot \vec{\sigma}_j~.
\end{equation}
Here, we take into account only its spin-spin component, which produces the most important
hyperfine forces for the baryon spectra. While $\vec{\sigma}_i$ represent the Pauli spin
matrices of $SU(2)_S$, the $\lambda_i^a$ are the generalized Gell-Mann flavor matrices of
$SU(5)_F$ for quark $i$. In addition
to the exchange of the pseudoscalar 24-plet also the flavor-singlet is included because of
the $U(1)$ anomaly. The radial form of the GBE potential resembles the one of the pseudoscalar
meson exchange
\begin{equation}
V_{\beta}(\vec{r}_{ij})=\frac{g_{\beta}^2}{4\pi}\frac{1}{12 m_i m_j}\bigg[ \mu_{\beta}^2 \frac{e^{-\mu_{\beta}r_{ij}}}{r_{ij}}-4\pi \delta(\vec{r}_{ij})\bigg]
\label{delta}
\end{equation}
for $\beta = 24$ and $\beta = 0$. Herein the $\delta$-function must be smeared out leading
to~\cite{Glozman:1997fs,Glozman:2000fu}
\begin{equation}
V_{\beta}(\vec{r}_{ij})=\frac{g_{\beta}^2}{4\pi}\frac{1}{12 m_i m_j}\bigg[ \mu_{\beta}^2 \frac{e^{-\mu_{\beta}r_{ij}}}{r_{ij}}- \Lambda_{\beta}^2 \frac{e^{-\Lambda_{\beta}r_{ij}}}{r_{ij}}\bigg] \,.
\end{equation}
Contrary to the earlier GBE RCQM~\cite{Glozman:1997ag}, which uses several different
exchange masses $\mu_\gamma$ and different cut-offs $\Lambda_\gamma$,
corresponding to $\gamma=\pi$, $K$, and $\eta$=$\eta_8$ mesons, we here managed to get
along with a universal GBE mass $\mu_{24}$ and a single cut-off $\Lambda_{24}$ for the 24-plet
of $SU(5)_F$. Only the singlet exchange comes with another mass $\mu_0$ and another cut-off
$\Lambda_0$ with a separate coupling constant $g_0$.
Consequently the number of open parameters in the hyperfine interaction could be kept as
low as only three (see Tab.~\ref{Free parameters}). 

\begin{table}[h]
\begin{ruledtabular}
\begin{tabular}{ccc}
\multicolumn{3}{c}{Free Parameters} \\
\hline
\\
$(g_0 / g_{24})^2$ &$\Lambda_{24}$ [fm$^{-1}$] &\vspace{1mm} $\Lambda_0$  [fm$^{-1}$] \\
\hline \\[-3mm]
1.5 & 3.55 & 7.52 \\
\end{tabular}
\end{ruledtabular}
\caption {Free parameters of the present GBE RCQM determined by a best fit to the baryon spectra.}
\label{Free parameters}
\end{table}

All other parameters entering the model have judiciously been predetermined by
existing phenomenological insights. In this way the constituent quark masses have been
set to the values as given in Tab.~\ref{Fixed parameters}. The 24-plet Goldstone-boson (GB)
mass has been assumed as the value of the $\pi$ mass and similarly the singlet mass as the
one of the $\eta'$. The universal coupling constant of the 24-plet has been chosen according
to the value derived from the $\pi$-$N$ coupling constant via the Goldberger-Treiman relation.

\begin{table}[h]
\begin{ruledtabular}
\begin{tabular}{ccccccc}
\multicolumn{7}{c}{Fixed Parameters}\\
\hline
%\vspace{0.1mm}\\
\\
\multicolumn{4}{c}{Quark masses [MeV]} & \multicolumn{2}{c}{Exchange masses [MeV]}& Coupling \\
$m_u=m_d$ & $m_s$ & $m_c$ & $m_b$ & \vspace{1mm} \hspace{3mm} $\mu_{24}$ & $\mu_{0}$ & $g^2_{24}/4\pi$ \\
\hline \\[-3mm]
340& 480 & 1675 & 5055 &  \hspace{3mm} 139 & 958 & 0.7 \\
\end{tabular}
\end{ruledtabular}
\caption {Fixed parameters of the present GBE RCQM predetermined from phenomenology and
not varied in the fitting procedure.}
\label{Fixed parameters}
\end{table}

%\section{Results}

\begin{figure}[t]
\includegraphics[width=7cm]{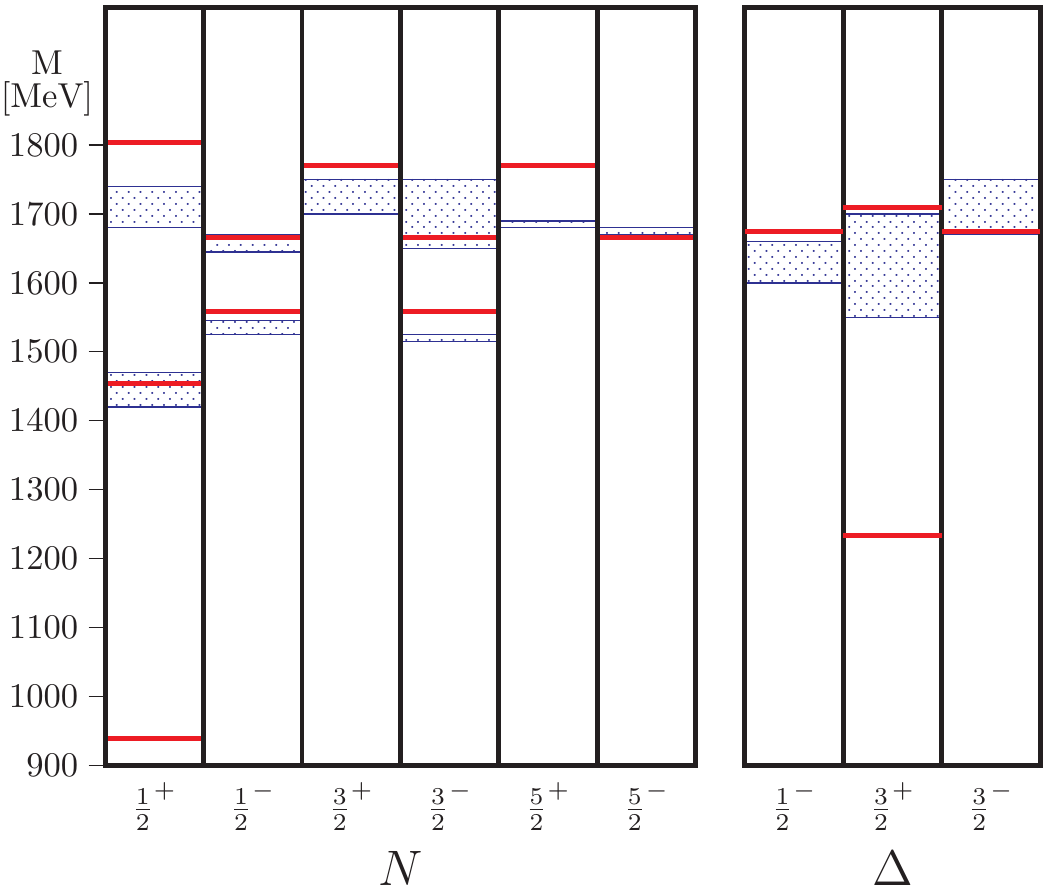}\\
\caption{\label{fig:light} Nucleon and $\Delta$ excitation spectra (solid/red levels)
as produced by the universal GBE RCQM in comparison to phenomenological
data~\cite{Nakamura:2010zzi}
(the gray/blue lines and shadowed/blue boxes show the masses and their uncertainties).}
\end{figure}

\begin{figure}[h]
\includegraphics[width=8.6cm]{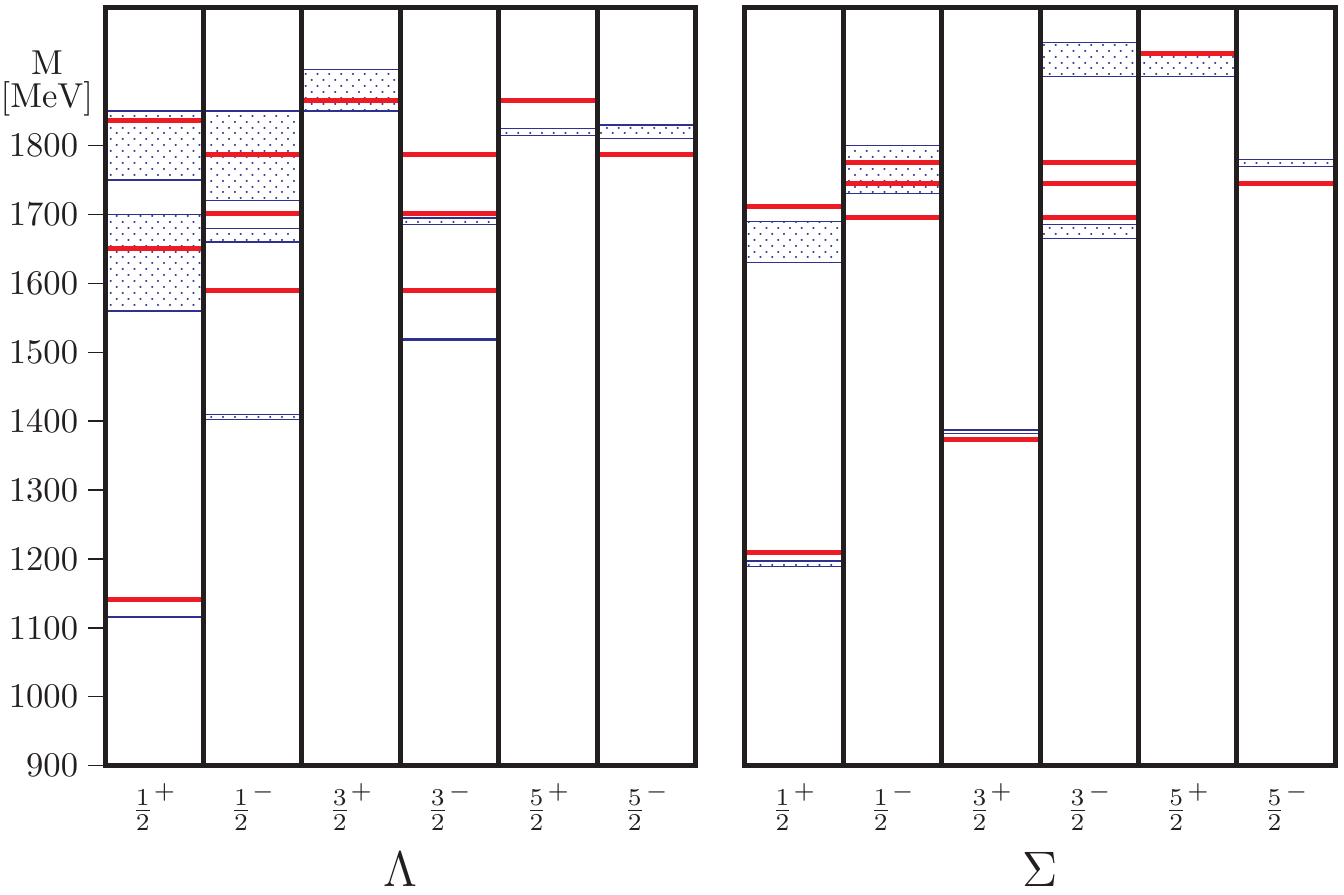}\\
\vspace{5mm}
\includegraphics[width=4.5cm]{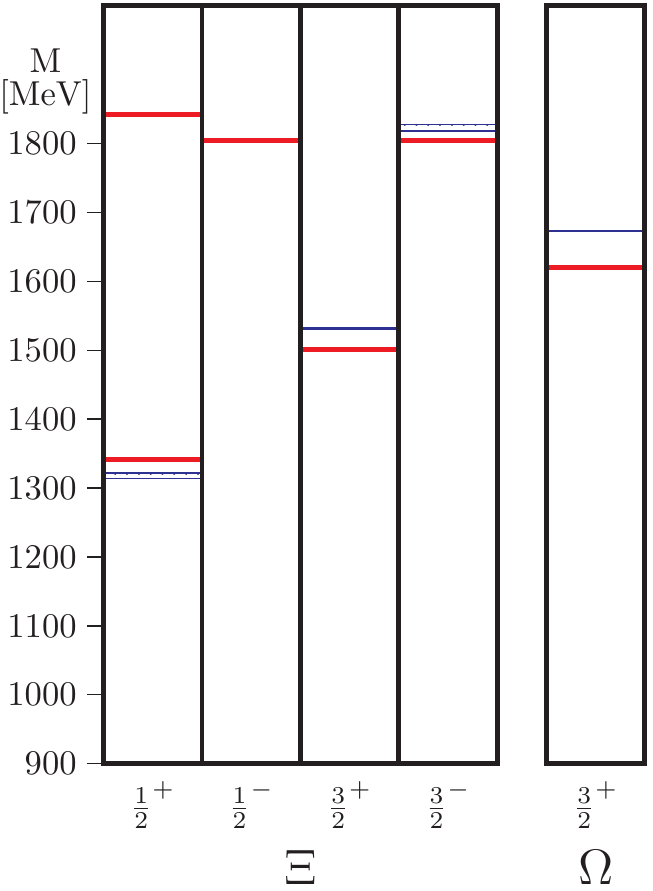}\\
\caption{\label{fig:strange} Same as Fig.~\ref{fig:light} but for the strange baryons.}
\end{figure}

We have calculated the baryon spectra of the relativistically invariant mass operator
$\hat M$ to a high accuracy both by the stochastic variational method~\cite{Suzuki:1998bn}
as well as the modified Faddeev integral equations~\cite{Papp:2000kp,McEwen:2010sv}.
The present universal GBE
RCQM produces the spectra in the light and strange sectors with
similar or even better quality than the previous GBE RCQM~\cite{Glozman:1997ag}
(see Figs.~\ref{fig:light} and~\ref{fig:strange}). Most importantly, the right level
orderings specifically in the $N$, $\Delta$ and $\Lambda$ spectra as well as all other
$SU(3)_F$ ground and excited states are reproduced in accordance with phenomenology.
The reasons are exactly the same as for the previous GBE RCQM, which has already been
extensively discussed in the literature~\cite{Glozman:1995fu,Glozman:1997ag,Glozman:1997fs}.
Unfortunately, the case of the $\Lambda$(1405) excitation could still not be resolved.
It remains as an intriguing problem for all three-quark CQMs.

\begin{figure}[t]
\includegraphics[width=8.6cm]{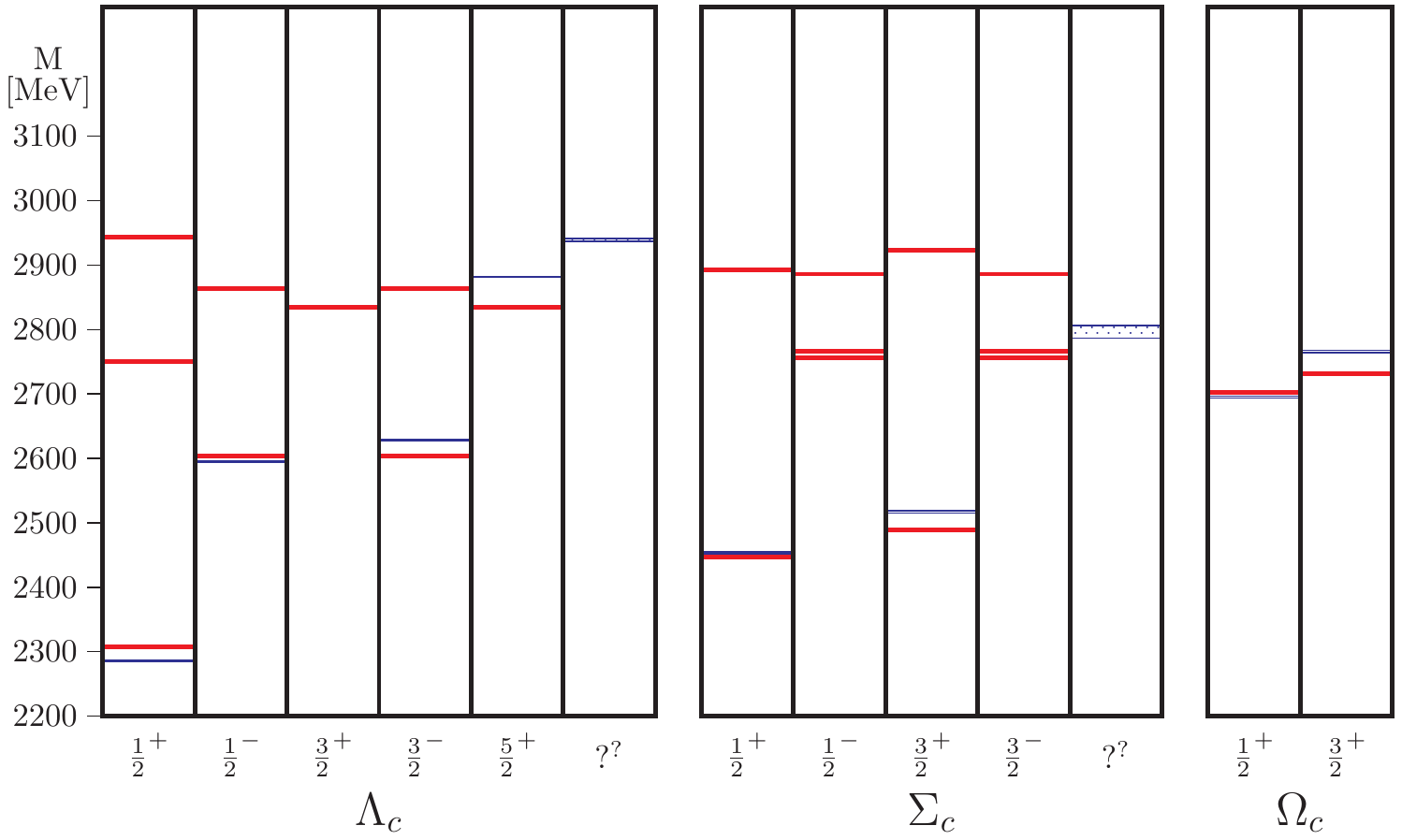}\\
\caption{\label{fig:charm}Same as Fig.~\ref{fig:light} but for charm baryons.}
\end{figure}

\begin{figure}[h]
\includegraphics[width=4cm,height=5cm]{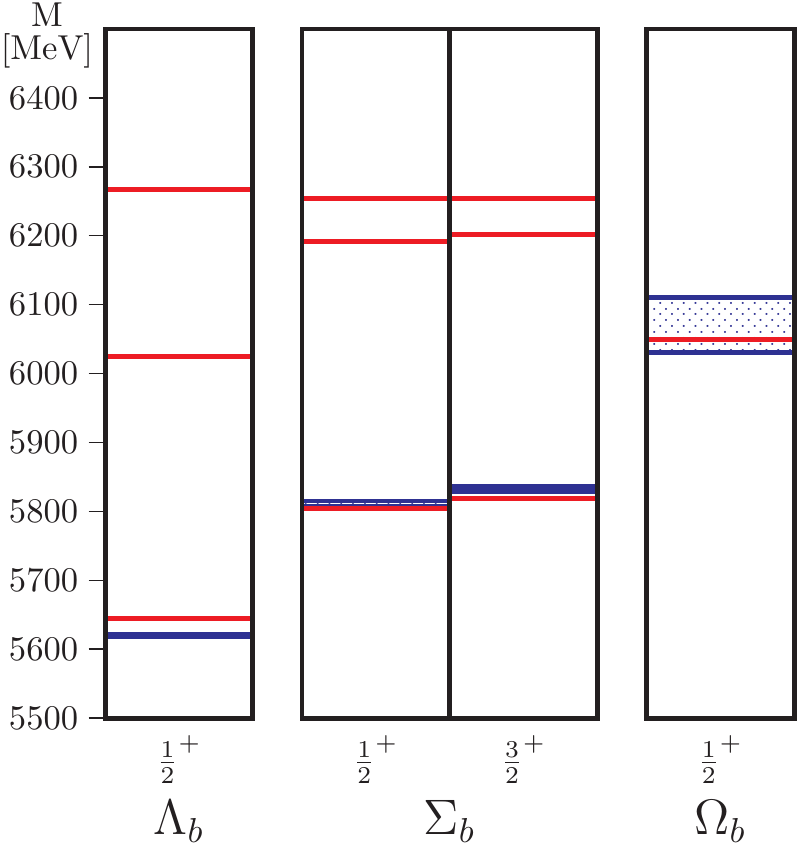}\\
\caption{\label{fig:bottom}Same as Fig.~\ref{fig:light} but for bottom baryons.}
\end{figure}

What is most interesting in the context of the present work are the very properties of the
light-heavy and heavy-heavy $Q$-$Q$ hyperfine interactions. Can the GBE dynamics reasonably
account for them? In Figs.~\ref{fig:charm} and~\ref{fig:bottom} we show the spectra
of all charm and bottom baryons that experimental data with at least three- or four-star status
by the PDG~\cite{Nakamura:2010zzi} are available for~\footnote{Only $\Xi_c$ and $\Xi_b$ 
are missing, as we are presently not in the position to calculate baryons
with three different constituent-quark masses.}.
As is clearly seen, our universal GBE RCQM can reproduce all levels with respectable accuracy.
In the $\Lambda_c$ and $\Sigma_c$ spectra some experimental levels are not known with
regard to their spin and parity $J^P$. They are shown in the right-most columns of
Fig.~\ref{fig:charm}. Obviously they could easily be accommodated in accordance with the
theoretical spectra, once their $J^P$'s are determined. Furthermore the model predicts some
additional excited states for charm and bottom baryons that are presently missing
in the phenomenological data base.

Of course, the presently available data base on charm and bottom baryon states is not yet
very rich and thus not particularly selective for tests of effective $Q$-$Q$ hyperfine
forces. The situation will certainly improve with the advent of further data from
ongoing and planned experiments. Beyond the comparison to experimental data, we note
that the theoretical spectra produced by our present GBE RCQM are also in good agreement
with existing lattice-QCD results for heavy-flavor baryons. This is especially
true for the charm baryons vis-\`a-vis the recent work by Liu et al.~\cite{Liu:2009jc}.
Further comparisons with results from lattice QCD and alternative methods will
be given in a forthcoming more detailed article~\cite{Day:2012}, where also a number
of additional theoretical spectra up to $\Omega_{bbb}$ will be presented (for which,
however, no phenomenological data exist so far).

%\section{Conclusion and Discussion}

We emphasize that the most important ingredients into the present RCQM are relativity,
or more generally Poincar\'e invariance, and a hyperfine interaction that is derived
from an interaction Lagrangian built from effective fermion (constituent quark) and boson
(Goldstone boson) fields connected
by a pseudoscalar coupling~\cite{Glozman:1995fu}. It appears that such kind of dynamics is
quite appropriate for constituent quarks of any flavor. The effects of the hyperfine
forces do not at all become tentatively small for baryons with charm and bottom flavors.
In some cases at least the heavy-light interactions are of the same importance for the level
spacings as the light-light interactions. This has already been seen for charm baryons
in the work by the Bonn group~\cite{Migura:2006ep} and is also true for our universal GBE
RCQM (as will be detailed in ref.~\cite{Day:2012} too). It is furthermore in
line with findings from earlier lattice-QCD calculations~\cite{Woloshyn:2000fe}.

As a result we have demonstrated by the proposed GBE RCQM that a universal description of
all known baryons is possible in a single model. Here, we have considered only
the baryon masses (eigenvalues of the invariant mass operator $\hat M$). Beyond spectroscopy
the present model will be subject to further tests with regard to the baryon eigenstates,
which are simultaneously obtained from the solution of the eigenvalue problem of $\hat M$.
They must prove reasonable in order to make the model a useful tool for the treatment of
all kinds of baryons reactions within a universal framework.

\section{Acknowledgement}
This work was supported by the Austrian Science Fund, FWF, through the Doctoral
Program on {\it Hadrons in Vacuum, Nuclei, and Stars} (FWF DK W1203-N16).

%\nocite{*}


\begin{thebibliography}{10}

\bibitem{Nakamura:2010zzi} 
  K.~Nakamura {\it et al.}  [Particle Data Group Collaboration],
  %``Review of particle physics,''
  J.\ Phys.\ G {\bf 37}, 075021 (2010)
  %%CITATION = JPHGB,G37,075021;%%

\bibitem{Loring:2001kx} 
  U.~L\"oring, B.C.~Metsch, and H.-R.~Petry,
  %``The Light baryon spectrum in a relativistic quark model with instanton induced quark forces: The Nonstrange baryon spectrum and ground states,''
  Eur.\ Phys.\ J.\ A {\bf 10}, 395 (2001); ibid. 447 (2001)
  
%\bibitem{Loring:2001ky} 
%  U.~L\"oring, B.~C.~Metsch and H.~R.~Petry,
%  %``The Light baryon spectrum in a relativistic quark model with instanton induced quark forces: The Strange baryon spectrum,''
%  Eur.\ Phys.\ J.\ A {\bf 10}, 447 (2001)
    
\bibitem{Migura:2006ep} 
  S.~Migura, D.~Merten, B.~Metsch, and H.-R.~Petry,
  %``Charmed baryons in a relativistic quark model,''
  Eur.\ Phys.\ J.\ A {\bf 28}, 41 (2006)

\bibitem{Semay:2007cv} 
  C.~Semay, F.~Buisseret, N.~Matagne, and F.~Stancu,
  %``Baryonic mass formula in large N(c) QCD versus quark model,''
  Phys.\ Rev.\ D {\bf 75}, 096001 (2007)

\bibitem{Semay:2007ff} 
  C.~Semay, F.~Buisseret, and F.~Stancu,
  %``Mass formula for strange baryons in large N(c) QCD versus quark model,''
  Phys.\ Rev.\ D {\bf 76}, 116005 (2007)

\bibitem{Semay:2008wn} 
  C.~Semay, F.~Buisseret, and F.~Stancu,
  %``Charm and bottom baryon masses in the combined $1/N_c$ and $1/m_Q$ expansion versus quark model,''
  Phys.\ Rev.\ D {\bf 78}, 076003 (2008)

\bibitem{SanchisAlepuz:2011aa} 
  H.~Sanchis-Alepuz, R.~Alkofer, G.~Eichmann, and R.~Williams,
  %``Model comparison of Delta and Omega masses in a covariant Faddeev approach,''
  PoS QCD-TNT-II, 041 (2011) [arXiv:1112.3214]

\bibitem{SanchisAlepuz:2012}
  H.~Sanchis-Alepuz, PhD Thesis, University of Graz (2012)  

\bibitem{Liu:2009jc} 
  L.~Liu, H.-W.~Lin, K.~Orginos, and A.~Walker-Loud,
  %``Singly and Doubly Charmed J=1/2 Baryon Spectrum from Lattice QCD,''
  Phys.\ Rev.\ D {\bf 81}, 094505 (2010)

\bibitem{Manohar:1983md} 
  A.~Manohar and H.~Georgi,
  %``Chiral Quarks and the Nonrelativistic Quark Model,''
  Nucl.\ Phys.\ B {\bf 234}, 189 (1984)
    
\bibitem{Glozman:1995fu} 
  L.Ya.~Glozman and D.O.~Riska,
  %``The Spectrum of the nucleons and the strange hyperons and chiral dynamics,''
  Phys.\ Rept.\  {\bf 268}, 263 (1996)

\bibitem{Weinberg:2010bq} 
  S.~Weinberg,
  %``Pions in Large-$N$ Quantum Chromodynamics,''
  Phys.\ Rev.\ Lett.\  {\bf 105}, 261601 (2010)

\bibitem{Glozman:1997ag} 
  L.Ya.~Glozman, W.~Plessas, K.~Varga, and R.F.~Wagenbrunn,
  %``Unified description of light and strange baryon spectra,''
  Phys.\ Rev.\ D {\bf 58}, 094030 (1998)

\bibitem{Glozman:1997fs} 
  L.Ya.~Glozman, Z.~Papp, W.~Plessas, K.~Varga, and R.F.~Wagenbrunn,
  %``Effective Q-Q interactions in constituent quark models,''
  Phys.\ Rev.\ C {\bf 57}, 3406 (1998) 

\bibitem{Boffi:2001zb} 
  S.~Boffi, L.Ya.~Glozman, W.~Klink, W.~Plessas, M.~Radici, and R.F.~Wagenbrunn,
  %``Covariant electroweak nucleon form-factors in a chiral constituent quark model,''
  Eur.\ Phys.\ J.\ A {\bf 14}, 17 (2002)

\bibitem{Plessas:2010pk} 
  W.~Plessas,
  %``Relativistic point-form approach to hadron properties,''
  PoS LC 2010, 017 (2010) [arXiv:1011.0156]

\bibitem{Glozman:1995xy} 
  L.Ya.~Glozman and D.~O.~Riska,
  %``The Charm and bottom hyperons and chiral dynamics,''
  Nucl.\ Phys.\ A {\bf 603}, 326 (1996)
  [Erratum ibid.\ A {\bf 620}, 510 (1997)]  
  
\bibitem{Glozman:2000fu} 
  L.Ya.~Glozman, Z.~Papp, W.~Plessas, K.~Varga, and R.~F.~Wagenbrunn,
  %``Reply to Comment on Effective Q-Q interactions in constituent quark models,''
  Phys.\ Rev.\ C {\bf 61}, 019804 (2000)

\bibitem{Suzuki:1998bn} 
  Y.~Suzuki and K.~Varga,
  {\it Stochastic Variational Approach to Quantum-Mechanical Few-Body Problems},
  Lect.\ Notes Phys.\ {\bf 54}, 1 (1998)

\bibitem{Papp:2000kp} 
  Z.~Papp, A.~Krassnigg, and W.~Plessas,
  %``Faddeev approach to confined three quark problems,''
  Phys.\ Rev.\ C {\bf 62}, 044004 (2000)  

\bibitem{McEwen:2010sv} 
  J.~McEwen, J.~Day, A.~Gonzalez, Z.~Papp, and W.~Plessas,
  %``Treatment of confinement in the Faddeev approach to three-quark problems,''
  Few-Body Syst.\  {\bf 47}, 225 (2010)
    
\bibitem{Day:2012} Joseph P.~Day, Ki-Seok Choi, and Willibald~Plessas, to be published  

\bibitem{Woloshyn:2000fe} 
  R.~M.~Woloshyn,
  %``Hyperfine effects in charmed baryons,''
  Phys.\ Lett.\ B {\bf 476}, 309 (2000)
                     
\end{thebibliography}
\end{document}